\documentclass[aps, prd, reprint, superscriptaddress, showpacs]{revtex4-1}
\usepackage{amsmath}
\usepackage{graphicx}
\usepackage{color}
\usepackage{subfigure}
\usepackage{booktabs}
\usepackage{siunitx}
\begin{document}

\title{Calculation of thermal noise in grating reflectors}
\author{D.~Heinert}
\email[]{daniel.heinert@uni-jena.de}
\affiliation{Institut f{\"u}r Festk{\"o}rperphysik, Friedrich-Schiller-Universit{\"a}t Jena, 07743 Jena, Germany}
\author{S.~Kroker}
\affiliation{Institut f{\"u}r Angewandte Physik, Friedrich-Schiller-Universit{\"a}t Jena, 07743 Jena, Germany}
\author{D.~Friedrich}
\affiliation{Institute for Cosmic Ray Research, The University of Tokyo, 5-1-5 Kashiwa-no-Ha, Kashiwa, Chiba 277-8582, Japan}
\author{S.~Hild}
\affiliation{SUPA, School of Physics and Astronomy, Institute for Gravitational Research, University of Glasgow, Glasgow G12 8QQ, United Kingdom}
\author{E.-B.~Kley}
\affiliation{Institut f{\"u}r Angewandte Physik, Friedrich-Schiller-Universit{\"a}t Jena, 07743 Jena, Germany}
\affiliation{Fraunhofer Institute of Applied Optics and Precision Engineering, Albert-Einstein-Stra{\ss}e 7, 07745 Jena, Germany}
\author{S.~Leavey}
\affiliation{SUPA, School of Physics and Astronomy, Institute for Gravitational Research, University of Glasgow, Glasgow G12 8QQ, United Kingdom}
\author{I.~W.~Martin}
\affiliation{SUPA, School of Physics and Astronomy, Institute for Gravitational Research, University of Glasgow, Glasgow G12 8QQ, United Kingdom}
\author{R.~Nawrodt}
\affiliation{Institut f{\"u}r Festk{\"o}rperphysik, Friedrich-Schiller-Universit{\"a}t Jena, 07743 Jena, Germany}
\author{A.~T{\"u}nnermann}
\affiliation{Institut f{\"u}r Angewandte Physik, Friedrich-Schiller-Universit{\"a}t Jena, 07743 Jena, Germany}
\affiliation{Fraunhofer Institute of Applied Optics and Precision Engineering, Albert-Einstein-Stra{\ss}e 7, 07745 Jena, Germany}
\author{S. P.~Vyatchanin}
\affiliation{Faculty of Physics, Moscow State University, Moscow 119991, Russia}
\author{K.~Yamamoto}
\affiliation{Institute for Cosmic Ray Research, The University of Tokyo, 5-1-5 Kashiwa-no-Ha, Kashiwa, Chiba 277-8582, Japan}

\begin{abstract}

Grating reflectors have been repeatedly discussed to improve the noise performance of metrological applications due to the reduction or absence of any coating material.
So far, however, no quantitative estimate on the thermal noise of these reflective structures exists.
In this work we present a theoretical calculation of a grating reflector's noise.
We further apply it to a proposed 3rd generation gravitational wave detector.
Depending on the grating geometry, the grating material and the temperature we obtain a thermal noise decrease by up to a factor of ten compared to conventional dielectric mirrors.
Thus the use of grating reflectors can substantially improve the noise performance in metrological applications.

\end{abstract}

\pacs{05.40.-a, 04.80.Nn,  42.79.Fm, 06.30.Ft}

\maketitle

%

\section{Introduction}

Current optical metrological experiments are operating at such high sensitivities that they can be crucially restricted by intrinsic noise processes. 
Among them the interferometric detection of gravitational waves as well as the frequency stabilization with optical cavities are limited by the thermal noise of their optical components \cite{hild2011,kessler2012}.
Up until now, the reflective optical components (mirrors,  cavity couplers,  etc.) of the above setups have been based upon the use of dielectric layer stacks (Bragg mirrors).
It turns out that the amorphous coatings with their high mechanical loss \cite{penn2003} represent the dominant thermal noise sources in these applications \cite{numata2004, Nawrodt2011}.
To overcome this limitation current research focusses on the coating materials.
Efforts targeted at understanding and optimizing the loss of current amorphous coating materials are ongoing \cite{bassiri2013,flaminio2010}, as are searches for alternative, low loss amorphous coating materials \cite{abernathy2011}.
Besides, crystalline coating layers as e.\,g. AlGaAs \cite{cole2008} are under investigation promising a low mechanical loss due to their crystalline structure.

Complementary approaches involve a conceptual revision of the reflective elements.
In this regard a potential alternative to Bragg mirrors can be found in the use of grating reflectors \cite{mashev1985, bunkowski2006}.
These periodic structures are known to exhibit reflection maxima at specific wavelengths. 
In contrast to Bragg mirrors, grating reflectors do not require several tens of alternating dielectric layers but rather only one structured high-index thin film embedded in an environment with a lower refractive index.  
Recently it has been theoretically and experimentally demonstrated that even monolithic crystalline silicon gratings can provide a high reflectance at a desired design wavelength. 
On a basic level there are two ways of implementation of these structures: free standing ridges (lamellar grating) \cite{zhou2009} or T-shape gratings \cite{brueckner2010}.
For the latter a reflectivity of 99.8\% has been experimentally demonstrated at \SI{1550}{\nano\meter}.  
They also provide the advantage of scalability to virtually arbitrary reflector areas.  
Silicon as the grating material further exhibits a low mechanical loss especially at cryogenic temperatures \cite{mcguigan1978, nawrodt2008}.
Thus, in contrast to amorphous materials it promises a further decrease of thermal noise by cooling.

With the reduced amount or absence of amorphous materials, a reduced thermal noise contribution is expected for grating reflectors. 
So far, however, no detailed and quantitative thermal noise analysis of such gratings exists.
Consequently, in the present work we model the different thermal noise contributions of grating reflectors. 
We further illustrate our model by an application to two specific grating configurations.
Finally, considering also the substrate noise contributions, the possible improvement in sensitivity with respect to thermal noise is presented for a gravitational wave detector as a potential metrological application.

\section{Basics on grating reflectors}

\begin{figure}[bt]
	\begin{center}
		\includegraphics[width=7cm]{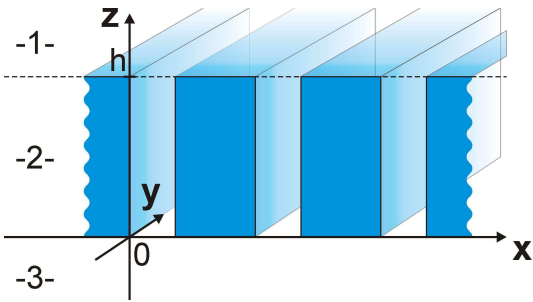}
		\caption{Sketch of a lamellar grating. A plane wave approaching the grating from the top (-1-) excites eigenmodes in the grating (-2-). A high reflectivity is obtained for a destructive interference at the output (-3-). This can occur effectively if the grating exhibits exactly two propagating modes.}
		\label{fig:lamellar}
	\end{center}
\end{figure}

The working principle of highly reflective gratings can be well understood on the simple geometry of a lamellar grating as has been investigated by Botten et al. \cite{botten1981}. 
They obtained the reflectivity spectrum of a lamellar grating surrounded by homogeneous media (see Fig.~\ref{fig:lamellar}) for plane wave incidence at the vacuum wavelength $\lambda_0$. 
Due to their increased tolerance to a wavelength or orientation mismatch we focus solely on the case of transversal-magnetic (TM) modes \cite{kroker2012b}.

The optical behavior of the grating is governed by the Helmholtz equation.
For the magnetic field $h_y$ it reads
\begin{align}
\Delta h_y+k_0^2n^2(x) h_y=\frac{\partial \ln \left[n^2(x)\right] }{\partial x}  \frac{\partial h_y}{\partial x}  \ ,
\label{equ:helmholtz}
\end{align}
with the wave vector $k_0=2\pi/\lambda_0$ and the refractive index $n$.
The most common solution scheme takes advantage of the problem's translational symmetry along the $x$-axis.
Due to Floquet's theorem the magnetic field can be decomposed with respect to the $x$-coordinate into
\begin{align}
h_y(x,z)=\sum_{s=-\infty}^\infty u_s(z) e^{\mathrm{i} 2\pi s x/\Lambda } \ ,
\end{align}
where $\Lambda$ represents the grating period and $u_s(z)$ the Fourier coefficients.
The decomposition above is simplified to the case of normal incidence.
It holds in any region of the grating and separates the spatial variables.
In the homogeneous regions (-1- and -3-) it is inserted into the Helmholtz equation (Eq.~(\ref{equ:helmholtz})) with a constant value of $n^2(x)$.
This yields uncoupled ordinary differential equations for each Fourier coefficient $u_s(z)$.
As these equations are of second order, two integration variables are introduced for each Fourier mode.
They can be understood as the amplitudes of an incoming and a reflected wave travelling along the $z$-direction.
A numerical calculation demands a truncation of the Fourier decomposition at the order $N$.
This results in the appearance of $2(2N+1)$ integration constants in each region (-1- and -3-).
In the physical problem half of them are known as they represent the incoming waves.
Consequently the homogeneous regions introduce $2(2N+1)$ unknowns in total.

In applying Eq.~(\ref{equ:helmholtz}) to the grating region -2- the term $n^2(x)$ is replaced by its decomposition in the Fourier space.
This leads to ordinary differential equations for $u_s(z)$ which in contrast to the homogeneous regions are coupled.
For a decomposition of order $N$ it yields $(2N+1)$ differential equations and $2(2N+1)$ unknowns.

To solve this problem a total of $4(2N+1)$ equations for the unkowns are necessary.
They can be found in the boundary conditions for the electric and the magnetic field at the borders between the grating regions. 
Here $h_y$ as well as the electric field component $e_x$ show a continuous behavior. 
Using Maxwell's equations the latter is revealed to be proportional to $\partial h_y/\partial z$.
On both boundaries these conditions result in $4(2N+1)$ equations and allow the computation of the transmitted and reflected light.
Most numerical calculations of gratings are based on this scheme known as RCWA (rigourously coupled wave analysis) \cite{moharam1981}. 
For a deeper mathematical treatment see e.\,g. Ref.~\cite{hench2008}.
As a well established and reliable numerical tool the RCWA method is used within this work to obtain the reflectance characteristics of possible grating reflectors.

In contrast to the numerical concept most analytical solutions of simple gratings follow a slightly different approach  while keeping the frame of calculation.
Namely they use grating eigenmodes $v_s(x)$ as the basic functions for a decomposition instead of Fourier components yielding
\begin{align}
h_y(x,z)=\sum_s v_s(x)e^{\mu_s z} \ .
\end{align}
The parameter $\mu_s$ can be calculated to an effective refractive index of the eigenmode via $n_{\mathrm{eff}}=\mathrm{i} \mu_s/k_0$.
These eigenmodes keep their field profile during a propagation along the $z-$direction.
Also, in contrast to the Fourier modes, they can be propagated independently through the grating region and thus allow a deeper physical insight into the grating properties.
Only a limited number of modes show real effective indices and thus a propagation along the $z$-direction. 
The other modes are evanescent and are not allowed to carry energy through the grating.

With this knowledge the effect of an incident plane wave on a lamellar grating is accessible.
The incident plane wave approaches from the region -1- of Fig.~\ref{fig:lamellar} along the negative $z$-direction. 
It encounters a first scatter at the top surface of the grating.
Here it is divided into transmitting eigenmodes of the grating (region -2-) as well as into reflected eigenmodes of the homogeneous medium (region -1-).
In this process evanescent modes are also excited.
The scattering matrix of the single surface is obtained due to the continuous behavior of the field $h_y$ at $z=h$.
Repeating this procedure at the bottom boundary at $z=0$ allows the characterization of the grating in terms of reflectivity.

The reflectivity of a lamellar grating is also obtainable by an analogy to a multi-mode Fabry-Perot resonator as suggested by Lalanne et al. \cite{lalanne2006}.
In this approach the evanescent modes are considered only in the calculation of the scattering matrix and neglected in the propagating terms. 
This approximation seems natural for deep gratings. 
The general idea of high reflection can then be understood by the following scheme. 
The grating geometry is chosen such that the incident plane wave excites two propagating modes in the grating. 
At the opposite boundary the grating modes are again partially transmitted and reflected.
In this process the energy redistribution between the grating modes is provided by the scattering matrix.
The consideration of repeated reflections then leads to the stationary mode intensity in the grating.
As both grating modes show different effective refractive indices along the $z$-direction, a change in the grating depth causes a relative phase shift between both propagating modes. 
In this way the transmitted parts can be modified to interfere destructively leading to highly reflective structures.
The interference of only two propagating modes in the grating also guarantees a large tolerance with respect to parameter deviations and a broadband reflective behavior.

\begin{figure}[tb]
	\begin{center}
		\includegraphics[width=7cm]{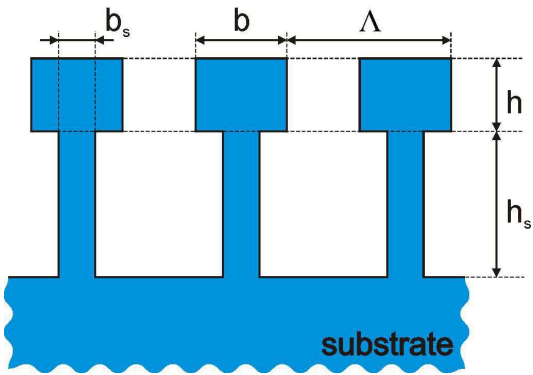}
		\caption{Geometry of a supported lamellar reflection grating with period $\Lambda$. 
		Both gratings are described by their depth $h$, $h_s$ and their width $b$, $b_s$. Further the fill factor $f$ is typically used to characterize the width $b$ via $b=f\Lambda$.
		In the topmost grating exactly two non-evanescent modes exist. The interference between them leads to the reflective behavior. The lower grating exhibits a single propagating mode and can be treated as a homogeneous medium with an effective refractive index.
		}
		\label{fig:geom}
	\end{center}
\end{figure}

For T-shaped gratings, basically the same scheme can be applied as they can be seen as being composed of two stacked lamellar gratings.
Their structure is presented in Fig.~\ref{fig:geom}.
The grating on top shows two optical modes and provides the high reflectivity and is therefore called the optical grating.
In contrast, the lower grating shows a decreased fill factor and is designed in a way to only exhibit a single propagating mode. 
With respect to its optical behavior it can therefore be treated as a homogeneous effective medium.
As it represents the mechanical support, we will refer to it as the supporting structure below.

\begin{table}[b]
\caption{Geometry parameters for a supported silicon grating used as the basic configuration in this paper. This grating is designed to show a maximum reflectivity at a wavelength of $\lambda_0=\SI{1550}{\nano\meter}$.}
\label{tab:supported}
\begin{ruledtabular}
\begin{tabular}{ll}
grating period $\Lambda$& \SI{688}{\nano\metre} \\
refractive index $n$&3.48 \\
\addlinespace[3mm]
\multicolumn{2}{l}{\textbf{optical grating}} \\
depth $h$& \SI{350}{\nano\metre} \\
fill factor $f=b/\Lambda$& \SI{.5643}{} \\
\addlinespace[3mm]
\multicolumn{2}{l}{\textbf{supporting structure}}\\
depth $h_s$ & \SI{800}{\nano\metre} \\
fill factor $f_s=b_s/\Lambda$& \SI{.25}{} \\
\end{tabular}
\end{ruledtabular}
\end{table}

\begin{table}[tb]
\caption{Geometry parameters of the underlying silicon substrate. The values are adopted to those proposed for ET-LF \cite{hild2011}.}
\label{tab:substrate}
\begin{ruledtabular}
\begin{tabular}{ll}
substrate material & silicon $\left\langle100\right\rangle$ \\
\addlinespace[3mm]
substrate radius $R$& \SI{.25}{\metre} \\
substrate height $H$& \SI{.46}{\metre} \\
beam radius $w_0=\sqrt{2}r_0$ &\SI{90}{\milli\metre}
\end{tabular}
\end{ruledtabular}
\end{table}

In this paper the scheme of a noise calculation is illustrated for a single fixed grating configuration. 
Its parameters are shown in Table~\ref{tab:supported}.
For this purpose we use a monolithic grating mirror made of silicon due to its low mechanical loss promising low noise levels.
The substrate dimensions are oriented at the low frequency detector of the Einstein Telescope (ET-LF) \cite{hild2011} and given in Table~\ref{tab:substrate}. 
The material properties of silicon are presented in Appendix~\ref{app:matprops}.
Using silicon in the optically active grating requires a low optical absorption to maintain high mirror reflectivities.
For this reason the grating has been optimized with respect to a maximum reflectivity at $\lambda_0=\SI{1550}{\nano\meter}$ by a numerical analysis using the RCWA method.
At this wavelength the photon energy is well below the fundamental band gap of silicon and promises an absorption sufficiently low for a use in ET-LF.
The grating geometry also fulfills the requirement of a technical fabrication in view of moderate grating depths and sufficiently large ridge widths of the supporting structure.
Due to the anisotropic etching process the fabrication of monolithic silicon gratings along the crystallographic $\left\langle 100\right \rangle$ or $\left\langle 110\right \rangle$-orientation is of particular interest.
In this work we assume a substrate orientation along the crystalline $\left\langle 100\right \rangle$-axis.
This orientation manifests in the mechanical analysis, i.\,e. in the choice of effective isotropic elastic parameters of silicon as an anisotropic material.

\section{Brownian grating noise}
\label{sec:BrownianNoise}
An interferometric gravitational wave detector is able to sense phase changes in the interferometer arms.
Thus any displacement of the reflecting surface of the test mass mirrors will introduce noise to the detector.
This noise is weighted by the intensity of the laser beam.

The calculation of Brownian noise is performed by using the direct approach by Levin \cite{levin1998}. 
At this point we shortly repeat Levin's scheme for a conventional mirror.
At first, a harmonic virtual pressure $p_\mathrm{L}(\vec{r})$ at an angular frequency $\omega$ is applied to the reflecting surface of the mirror.
It is weighted by the intensity distribution of the laser beam, i.\,e. the radially symmetric fundamental Gaussian mode, and reads
\begin{align}
p_\mathrm{L}(r)=\frac{F_0}{\pi r_0^2}\exp \left(-\frac{r^2}{r_0^2}\right) \ .
\end{align}
Here $F_0$ is an arbitrary constant characterizing the total applied force to the surface and $r_0$ represents the radius where the laser intensity has dropped to 1/e of its maximum value.
In the second step the rate of energy dissipation $P_\mathrm{diss}$ has to be calculated with respect to the virtual load.
With the elastic energy density $w_\mathrm{el}$ and the mechanical loss $\phi$ the model of structural damping \cite{levin1998} yields
\begin{align}
P_\mathrm{diss}=\omega \int_V \phi(\vec{r}) w_\mathrm{el}(\vec{r}) dV \ ,
\label{equ:structdamp}
\end{align}
where the integration is performed over the sample volume $V$.
Finally the noise power spectrum at frequency $\omega$ is calculated via
\begin{align}
S_z(\omega)=\frac{8k_\mathrm{B}T}{\omega^2}\frac{P_\mathrm{diss}}{F_0^2} \ .
\label{equ:Sxlevin}
\end{align}
Here the Boltzmann constant $k_\mathrm{B}$ and the temperature $T$ enter the expression.

Although the main scheme of a noise calculation remains the same, the detailed application changes for a grating reflector.
As the latter does not show a single reflecting surface but ridges also the application of forces becomes inhomogeneous.
Any displacement of a grating ridge may affect the phase of the reflected light and thus can introduce thermal noise to the setup.
This phase noise is sensed by any interferometric setup such as e.\,g. gravitational wave detectors and leads to a noise signal at the detector's output.
On the other hand, the gratings are operated at maximum reflectivity, so the amplitude fluctuations of the reflected light have to vanish in linear order.

\begin{figure}[bt]
	\begin{center}
		\subfigure[]
		{   	\includegraphics[height =4cm] {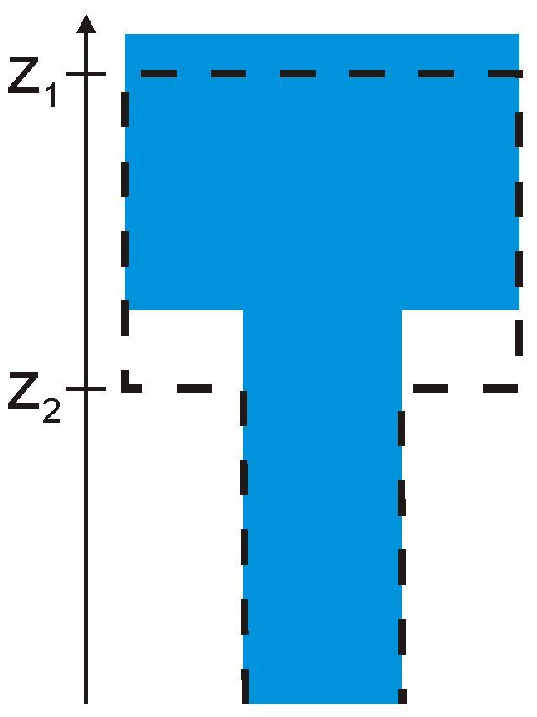}
   				\label{fig:disp_height}
 		}
		\subfigure[]
		{   	\includegraphics[height =4cm] {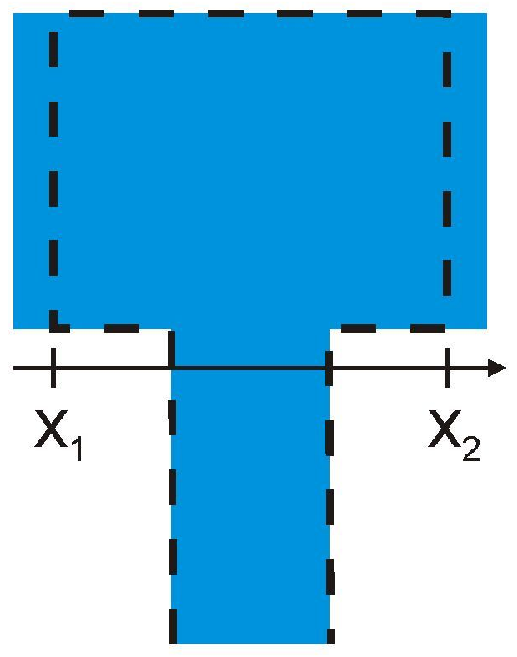}
   				\label{fig:disp_width}
 		}
		\caption{Illustration of the types of displacement leading to a phase change in the reflected light. A fluctuation in the grating height (a) as well as in the grating width (b) will influence the phase of the reflected light. The unperturbed grating geometry is indicated by the dashed line. The incoming light approaches from the top of the sketch.}
		\label{fig:disp}
	\end{center}
\end{figure}

In this work we distinguish displacements with respect to a change in the ridge height as well as in the ridge width as shown in Fig.~\ref{fig:disp}.
For the latter only a symmetric expansion has been considered. 
Due to the symmetry of the problem a pure translation of a single ridge along the grating plane must show a vanishing effect on the reflected light's phase in the linear order of displacement $\delta x$. 
Consequently, a pure translation is neglected in the following noise analysis.

The respective variation in the grating parameters will lead to different path lengths for the grating modes and cause a phase shift in the reflected light.
As the grating period is small compared to the radius of the laser beam $\Lambda\ll r_0$ the incoming laser light can be approximated as a plane wave with constant amplitude along a few ridges.
In the first step the effect of any ridge displacement on the phase of the reflected wave has to be calculated for an incoming plane wave.
This is a standard task for the RCWA method \cite{moharam1981, hench2008} which is used throughout this paper to compute the optical behavior of the grating reflector.

In the following we consider the effect of a homogeneous displacement of the grating ridges.
At first the impact of a fluctuation along the $z$-direction (see Fig.~\ref{fig:disp_height}) on the phase of the reflected light is investigated.
With $\delta z_1$ and $\delta z_2$ as the displacements of the front and rear interface of the optical grating, respectively, it reads
\begin{align}
\delta \varphi  &=-2k_0 \delta z_1+\frac{d\varphi_G}{dh}(\delta z_1-\delta z_2) \\
								&=\left(-2k_0+ \frac{d\varphi_G}{dh}\right)\delta z_1-\frac{d\varphi_G}{dh} \delta z_2 \ . \label{equ:dphi}
\end{align}
The expression $d\varphi_G/dh$ is characteristic for the particular grating and can be calculated numerically by RCWA.
Our grating geometry reveals a value of $d\varphi_G/dh=\SI{8.26e6}{\per\meter}$.
This value has been checked with an analytical approach adapting the method of Karagodsky et al. \cite{karagodsky2010} to the case of a T-shaped grating.
With a deviation of better than 1\% the analytical model confirms the results of the RCWA code.

Eq.~(\ref{equ:dphi}) can be rewritten in the following form
\begin{align}
\delta \varphi  &=2k_0 (e_1\delta z_1+e_2 \delta z_2) \ ,
\end{align}
with the weighting factors
\begin{align}
e_1&=-1+\frac{1}{2k_0}\frac{d\varphi_G}{dh}=0.019  \ , \\ 
e_2&=-\frac{1}{2k_0}\frac{d\varphi_G}{dh}=-1.019 \ .
\end{align}
These weighting factors $e_i$ determine the distribution of the virtual load at each interface.
The values above indicate that the light mainly probes a displacement of the rear surface $\delta z_2$.
Consequently, the virtual force is to be mainly applied to the rear surface. 
With respect to the different signs the two virtual forces at the optical grating have to be applied in an antiparallel orientation.

The unequal distribution of the virtual loads might seem counter-intutitive but can be qualitatively illustrated as follows.
Assume a single deformation of the front surface $\delta z_1>0$.
Here a part of the incoming wave is directly reflected.
This light shows a negative phase shift.
The remaining light couples into the grating region, propagates through the grating, is reflected and leaves the grating again.
Here the geometrical pathlength is increased by $2\delta z_1$.
Depending on the effective refractive index of the grating eigenmodes this change can lead to a positive phase change.
The interference with the first part can thus lead to a compensation resulting in a small value $e_1$.
Assuming a deformation at the rear surface $\delta z_2>0$ the situation differs.
Here the first, directly reflected part shows no phase change.
But the second contribution circulating through the grating still exhibits the same phase change as before.
Consequently, the resulting phase change of the reflected light is increased leading to the high value of $e_2$.

In a second analysis the effect of a width fluctuation of the optical ridges is estimated.
The phase change in the reflected light reads (c.\,f. Fig.~\ref{fig:disp_width})
\begin{align}
\delta \varphi  &=\frac{d\varphi_G}{db} (\delta x_2-\delta x_1)\ ,							
\end{align}
with $d\varphi_G/db=\SI{4.07e6}{\per\meter}$.
As only a symmetric deformation introduces a phase noise both displacements are assumed to equal each other. 
Then the weighting factor for both sides can be introduced as
\begin{align}
e=e_{x2}=-e_{x1}=\frac{1}{2k_0}\frac{d\varphi_G}{db}=0.50 \ .
\end{align}
In terms of a direct noise analysis this represents a symmetric compression of the optical grating ridge.

Besides the optical grating the reflected light also probes the supporting structure.
Consequently, a fluctuation in this region will lead to a phase noise in the reflected light.
A homogeneous change of the thickness $b_s$ results in the coefficient $d\varphi_G/db_s=\SI{5.14e6}{\per\meter}$.
This gives a value for the total force applied to the supporting structure as $e_s=0.63$.
In contrast to the mechanisms mentioned above the effect of a supporting width fluctuation will strongly depend on the $z$-position.
A fluctuation at the upper end will especially cause a significant effect.
For a quantitative estimate we computed the effect on the reflected light with respect to the $z$-dependence via RCWA. 
The results are presented at the end of this section.

\begin{figure}[tb]
	\begin{center}
		\includegraphics[height=4cm]{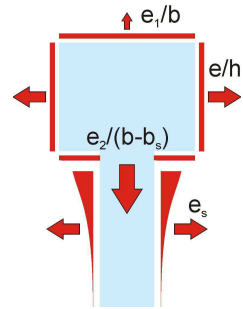}
		\caption{Application sketch of virtual pressures on the grating structure. The expressions characterize the strength of the applied virtual pressures. The elastic decoupling of the ridges allows an independent calculation of the elastic energy in each ridge.}
		\label{fig:load}
	\end{center}
\end{figure}

From this knowledge of the optical response of the grating to a ridge displacement, virtual forces can be deduced.
As described in Appendix~\ref{app:levinscheme}, these virtual loads have to be applied only to the air-grating-boundaries in a way that the total force equals the weighting parameters presented above.
This results in the load scheme shown in Fig.~\ref{fig:load}.

Following Levin's direct noise calculation the total elastic energy in the grating has to be determined in the next step.
For the calculation of elastic energy in a ridge we use a plane strain analysis.
This model is motivated by the fact that the intensity profile of the laser is typically only modulated over the beam radius $r_0$ on a scale of several centimeters.
Along this distance the virtual pressures on the ridge are also constant.
A plane strain analysis is then justified by the fact that this distance is large compared to the dimensions of the ridge's cross section.
Details of the plane strain calculation are presented in Appendix~\ref{app:plainstrain} for a simplified homogeneous load scheme. 

To clearly distinguish the effect of the different deformations, we treat them independently.
This procedure allows the identification of the most critical displacement.
To further obtain an analytical solution, we introduce the simplifying assumption that all probing forces are distributed homogeneously along the respective surface.
The validity of this assumption is evaluated later by a comparison to a result derived by a finite element analysis (FEA).  

We start with the probing forces along the $z$-direction. 
The loads weighted by $e_1$ and $e_2$ can be decomposed into a symmetric compression (due to $e_1$) and a one-sided application of a single force at the rear interface.
The elastic energy of the optical ridges due to the symmetric compression $e_1$ is calculated following Eq.~(\ref{equ:Etotplanestrain}) and reads
\begin{align}
E_\mathrm{tot}&=\frac{1-\sigma^2}{4Y} \frac{h}{\pi r_0^2} \frac{1}{f} e_1^2F_0^2\ .
\label{equ:Etot}
\end{align}
Here $Y$ and $\sigma$ represent the Young's modulus and Poisson's ratio of the grating material and $f$ the fill factor of the grating as given in Table~\ref{tab:supported}.

With the help of Eqs.~(\ref{equ:structdamp}) and (\ref{equ:Sxlevin}) the noise due to compression for a homogeneously distributed loss is accessible as
\begin{align}
S_z(\omega)=\frac{2}{\pi}\frac{k_BT}{\omega}\frac{h}{r_0^2 f}\frac{1-\sigma^2}{Y} e_1^2 \phi  \ ,
\end{align}
with the mechanical loss $\phi$ of the grating material.
This noise expression only considers the energy stored in the $z$-components of the elastic fields.
In terms of the analytical treatment of a multilayer coating by Harry et al. \cite{harry2002} it is called the perpendicular contribution of grating noise. 
Another contribution originates from the bending of the ridges due to the deformation of the substrate.
A detailed analysis of this remaining parallel energy contribution is shown in Appendix~\ref{app:bending}.
It reveals a negligibly small effect and thus is omitted in the rest of this work.

The remaining non-symmetric load characterized by $e_1+e_2$ equals 1 and represents a force at the rear side of the grating.
This force will introduce energy into the supporting structure via compression and consequently will cause thermal noise.
Although the load situation is different to the previous case, the stress in the supporting grating is constant due to the small dimensions compared to the substrate.
Thus the same scheme can be used again.
In analogy to Eq.~(\ref{equ:Etot}) we find a total elastic energy of
\begin{align}
E_{tot}=\frac{1-\sigma^2}{4Y} \frac{h_s}{\pi r_0^2}\frac{1}{f_s} F_0^2 \ ,
\end{align}
revealing a noise level of
\begin{align}
S_z(\omega)=\frac{2}{\pi}\frac{k_BT}{\omega}\frac{h_s}{r_0^2 f_s}\frac{1-\sigma^2}{Y} \phi \ .
\end{align}
Due to the increased grating depth, the decreased fill factor and the increased weighting factor, the noise amplitude from the supporting structure exceeds the contribution from the optical grating by the factor
\begin{align}
\sqrt{\frac{h_s}{h} \frac{f}{f_s} \frac{1}{e_1^2}}\approx 120 \ .
\end{align}

As a second step, the noise due to lateral displacements ($x$-direction) is investigated.
In this case the force is applied along the $x$-direction.
Instead of the top surface now the displacement on the ridge's side is responsible for a phase fluctuation.
Consequently, in the equations above the roles of ridge width $b$ and height $h$ are to be interchanged, e.\,g. the fill factor $f=b/\Lambda$ should be replaced by $h/\Lambda$. 
It turns out that this is the only modification required to adapt the above equations to width fluctuations.
The noise spectrum due to width fluctuations of the optical grating then reads
\begin{align}
S_z(\omega)=\frac{2}{\pi}\frac{k_BT}{\omega}\frac{b\Lambda}{h}\frac{1}{r_0^2}\frac{1-\sigma^2}{Y} e^2 \phi \ .
\end{align}
Compared to the height fluctuations of the supporting structure the noise amplitude due to width fluctuations of the optical grating is lower by the factor
\begin{align}
\sqrt{\frac{h}{b} \frac{h_s}{b_s}\frac{1}{e^2}}\approx 4 \ .
\end{align}
The calculation scheme can be repeated for the width fluctuations of the supporting structure.
For a homogeneous readout due to the increased height and the decreased width the respective noise power will be a factor of 3 lower compared to the optical grating.
In summary the height fluctuations of the supporting structure turn out to dominate the Brownian noise of our grating geometry for the simplified case of homogeneous loads.

To evaluate the strain energy density with respect to the real load scheme of Fig.~\ref{fig:load} an FE analysis has been performed.
There a single grating ridge on top of the substrate has been investigated numerically using a plane strain model.
Such a treatment is justified by the small elastic correlation between neighbored ridges.
At first only the homogeneously loaded supporting structure was investigated.
As a measure for comparison a modified linear energy density was calculated for the height compression of the supporting structure.
From Eq.~(\ref{equ:linEdens}) we obtain
\begin{align}
\pi^2 r_0^4 E_{y}&=b_s h_s\frac{1-\sigma^2}{2Y}\left(\frac{\Lambda}{b_s} \right)^2 F_0^2 \\
&=\SI{7.80e-24}{\square\meter\per\newton} \times F_0^2 \ ,
\end{align}
with the linear energy density $E_{y}$ of the supporting structure. 
The equivalent numerical calculation shows an agreement to better than 1 per thousand and demonstrates the accuracy of the FEA.
In a second numerical calculation we considered the complete grating with the real load case but with the simplification of a homogeneously distributed pressure at the side of the supporting structure.
This results in a factor of \SI{10.9e-24}{\square\meter\per\newton} showing an increase by 40\%.
A highly reflecting grating shows an exponential decay of light due to the evanescent character of its modes in the support.
Consequently a third FE calculation with an exponentially distributed force at the side of the supporting structure was performed.
For this purpose an RCWA calculation revealed a typical optical decay length of \SI{12}{\nano\meter}.
The inhomogeneous readout effectively reduces the averaging of ridge displacements. 
It is thus expected to exhibit an increased noise level.
This consideration has been confirmed by the mechanical FE calculation.
The resulting value of \SI{14.0e-24}{\square\meter\per\newton} indicates a total increase by 80\%.
To obtain an upper bound on the expected noise the last value was used to calculate the noise levels shown in Sec.~\ref{sec:results}.

Nevertheless, the analytical approach presented in this work is suited to obtain the correct order of magnitude of thermal noise.
Further, it is a valuable tool to identify the key parameters limiting the noise performance.
This also enables an effective optimization of grating reflectors with respect to a minimum thermal noise.


\section{Noise due to temperature fluctuations}
Besides the volume fluctuations also fluctuations in temperature form a source term of thermal noise.
The latter cause a phase noise in the reflected beam via thermal expansion $\alpha$ and the thermooptic effect characterized by $\beta=dn/dT$ .
These mechanisms are called thermoelastic \cite{braginsky1999} and thermorefractive \cite{braginsky2000} noise, respectively.

The temperature fluctuations as the source of this noise can be characterized by the thermal path length 
\begin{align}
r_{th}=\sqrt{\kappa/ (\rho C  \omega)} \ ,
\end{align}
with thermal conductivity $\kappa$, specific heat $C$, mass density $\rho$ and the radial frequency of fluctuations $\omega$.
Along this distance the temperature profile is approximately constant.
Only at larger scales the temperature can change significantly. 
For silicon at room temperature and a frequency of $f=\SI{100}{\hertz}$, typical for the detection band of gravitational wave detectors, a value of $r_{th}=\SI{380}{\micro\meter}$ is found.
As this value is two orders of magnitude larger than the grating period $\Lambda$ and the grating depth $h+h_s$ the fluctuating temperature can be assumed as constant along several hundred grating ridges.
Consequently, in a noise calculation a homogeneous temperature change can be assumed within a few ridges.
A decrease in temperature further increases the thermal path length and so further improves the reliability of this approximation.

Bearing this information in mind the calculation scheme of Braginsky et al. \cite{braginsky2003} for the temperature fluctuations in a small surface layer can be fully adopted to the case of grating reflectors.
They present the noise power of temperature fluctuations as
\begin{align}
S_{T}(\omega)=\frac{\sqrt{2}k_\mathrm{B}T^2}{\pi r_0^2\sqrt{\rho C\kappa \omega}} \ .
\label{equ:ST}
\end{align}
Below the effect of a homogeneous temperature change on the reflected light's phase is determined with the help of the RCWA code.
With this knowledge the phase noise of the reflected light becomes accessible.

\subsection{Thermorefractive noise}
Thermorefractive (TR) noise is caused by thermal fluctuations of the refractive index.
Thus, the effect of a refractive index change onto the phase of the reflected light has to be known for a grating reflector.
On the one hand such a variation in the optical grating will lead to a change of the effective indices of the optical grating modes.
Consequently, it causes a phase shift in the reflected light. 
On the other hand there is also an alteration in the electromagnetic field profile of each grating mode.
This leads to a variation of the scattering matrix at the grating boundaries and also modifies the reflected light's phase.
In the linear approximation of small changes a factor $K_\mathrm{TR}$ will appear, that links the phase change $\Delta \varphi$ to the refractive index change $\Delta n$ as
\begin{align}
\Delta \varphi=K_\mathrm{TR}\Delta n \ .
\end{align}
Using an RCWA code both effects are determined numerically for the presented grating configuration as
\begin{align}
K_\mathrm{TR,opt}=1.16 \ .
\end{align}

At first glance it seems counter-intuitive that a refractive index change in the support grating will also show an effect on the reflected light.
But even if no light propagates in the supporting structure, a change in its grating mode shapes will again alter the scatter behavior at the boundary and, thus, induce a phase change. 
Numerically we obtain
\begin{align}
K_\mathrm{TR,supp}=0.21\ .
\end{align}
In a coherent treatment of the complete structure a total TR coefficient
\begin{align}
K_\mathrm{TR}=K_\mathrm{TR,opt}+K_\mathrm{TR,supp}=1.37 \ ,
\end{align}
is obtained.

The final noise result is stated in terms of an effective test mass displacement 
\begin{align}
\Delta z=\frac{\lambda_0}{4\pi} \Delta \varphi \ .
\end{align}
The noise power with respect to this effective displacement can then be expressed via
\begin{align}
S_z(\omega)=\left(\frac{\lambda_0}{4\pi}K_\mathrm{TR}\beta \right)^2 S_T(\omega) \ .
\end{align}

\subsection{Thermoelastic noise}
Thermoelastic (TE) noise results from the thermal expansion $\alpha$ of the grating structure.
A homogeneous temperature change $\Delta T$ will lead to a homogeneous length change of the structure keeping the aspect ratio of the ridge constant.
Here $\delta=\alpha \Delta T $ characterizes the relative size deviations.
The homogeneous size change will contribute to a phase shift as
\begin{align}
\Delta \varphi=K_\mathrm{TE} \delta \ .
\end{align}
Using the RCWA code the constant $K_\mathrm{TE}$ was calculated numerically.

We again divide our investigation into contributions from the optical and the support grating.
At first the optical grating is considered.
Keeping the substrate surface fixed a thermal expansion of the optical grating is described by
\begin{align}
K_\mathrm{TE,opt}=1.62 \ .
\end{align}
Further also a variation in the supporting structure's width will influence the reflected phase via the modification of the optical mode shapes (as described for TR noise).
An analysis of this part leads to 
\begin{align}
K_\mathrm{TE,supp}=0.89 \ .
\end{align}
A variation of the support grating's depth won't change the mode shape.
Additionally the depth $h_s$ is also chosen to effectively decouple evanescent grating modes from the substrate.
Such a coupling would limit the maximum reflectivity of a grating reflector.
Nevertheless for our exemplaric grating geometry this effect has been identified to lead to a transmittance well below $10^{-6}$ at the working point and can be neglected.
From these considerations no effect is expected for a change in the support grating's depth.

But due to the massive substrate the expansion of the support will lead to a motion of the optical grating towards the beam and thus induces a phase change.
This effect reads
\begin{align}
\Delta \varphi &=-\frac{4\pi}{\lambda_0} \, h_s \, \alpha \Delta T \\
								&=K_{\mathrm{TE,}l} \delta \ .
\end{align}
The last equation defines the coefficient $K_{\mathrm{TE,}l}$.
For our parameters a value of $K_{\mathrm{TE,}l}=-6.49$ is found.

A temperature fluctuation at the substrate's surface further leads to a change in the grating period.
The numerical analysis of this effect on the reflected light reveals a last coefficient of 
\begin{align}
K_{\mathrm{TE,}p}=3.23 \ .
\end{align}

Finally the TE noise power spectrum can be calculated by the coefficient
\begin{align}
K_\mathrm{TE}&=K_\mathrm{TE,opt}+K_\mathrm{TE,supp}+K_{\mathrm{TE,}l}+K_{\mathrm{TE,}p}\\
\nonumber &=-0.75 \ ,
\end{align}
via
\begin{align}
S_z(\omega)=\left(\frac{\lambda_0}{4\pi}K_\mathrm{TE}\alpha \right)^2 S_T(\omega) \ .
\end{align}

\subsection{Thermooptic noise}
As TR and TE noise are driven by the same source, namely temperature fluctuations, they are correlated.
The coherent noise named thermooptic (TO) can be effectively decreased compared to an independent summation of the single contributions \cite{evans2008, gorodetsky2008}.
The TO noise power is then characterized by
\begin{align}
S_z(\omega)=\left(\frac{\lambda_0}{4\pi}\right)^2  \left(K_\mathrm{TR} \beta+K_\mathrm{TE}\alpha \right)^2 S_T(\omega) \ .
\end{align}

For our geometry TR noise outnumbers TE noise and thus prevents an effective compensation.
Further, due to the sign of the $K$-parameters, only the TE length change of the support grating allows a noise reduction in terms of a TO treatment.
In contrast to Bragg mirrors a complete TO compensation is possible in grating reflectors.
For this purpose an increase of the supporting grating depth $h_s$ by a factor $\sim 15$ is necessary.
This in turn would largely increase Brownian noise spoiling the grating performance.
Further, it is a challenging demand on technological feasibility, too.
For these reasons the coherent TO treatment is omitted in this paper.

%
\section{Results}
\label{sec:results}
\subsection{Monolithic silicon grating}
In Table~\ref{tab:noiseamps} we summarize our results for the different noise contributions of the presented grating at a frequency of \SI{100}{\hertz} and for \SI{300}{\kelvin} and \SI{10}{\kelvin}.
There we also present the Brownian noise level for a homogeneous silicon layer of the same height $h+h_s$ as the grating.
For this purpose the theory of Harry et al. \cite{harry2002} was used.
Interestingly, this approximation deviates from the precise results. 
For our grating geometry the noise level is underestimated by a factor of 2.
This result can be explained by the following rough considerations.
Firstly the incoming light mainly probes the fluctuations of the supporting structure.
Compared to an usual coating the respective virtual pressure in a grating is larger by the inverse fill factor $1/f_s$ as given in Eq.~(\ref{equ:levinpress}).
Consequently the strain energy density in the grating is larger by a factor of $1/f_s^2$.
To obtain the total elastic energy an integration along the $x$-axis introduces a multiplication by the fill factor.
Finally, the square root of the total elastic energy enters the noise amplitude.
It is proportional to $\sqrt{1/f_s}$ and reproduces the observed factor of two for a fill factor $f_s=0.25$ assumed in this work.

\begin{table}[bt]
\caption{Noise amplitude of the monolithic silicon grating discussed throughout this paper at a frequency of \SI{100}{\hertz}. The grating Brownian noise level is compared to the result of a single effective layer by the theory of Harry et al. \cite{harry2002}. The underlying material properties of silicon are presented in Table~\ref{tab:siprops}.}
\label{tab:noiseamps}
\begin{ruledtabular}
\begin{tabular}{lcc}
																&		$T=\SI{300}{\kelvin}$	&	$T=\SI{10}{\kelvin}$ \\
\hline
{\textbf{Noise amplitude} [\SI{}{\meter\per\sqrt{\hertz}}]} \\
Brownian grating														& \SI{1.45e-21}{}		& \SI{1.19e-22}{}\\
effective layer \cite{harry2002}						& \SI{0.72e-21}{}		& \SI{0.59e-22}{} \\
TE grating																	&	\SI{4.54e-24}{}		& \SI{1.06e-28}{}\\
TR grating																	& \SI{5.70e-22}{}  	& \SI{3.87e-26}{}\\					
\end{tabular}
\end{ruledtabular}
\end{table}

\begin{figure}[tbp]
  \centering
  \subfigure[\ $T=\SI{300}{\kelvin}$]{
    \includegraphics[width=7.5cm]{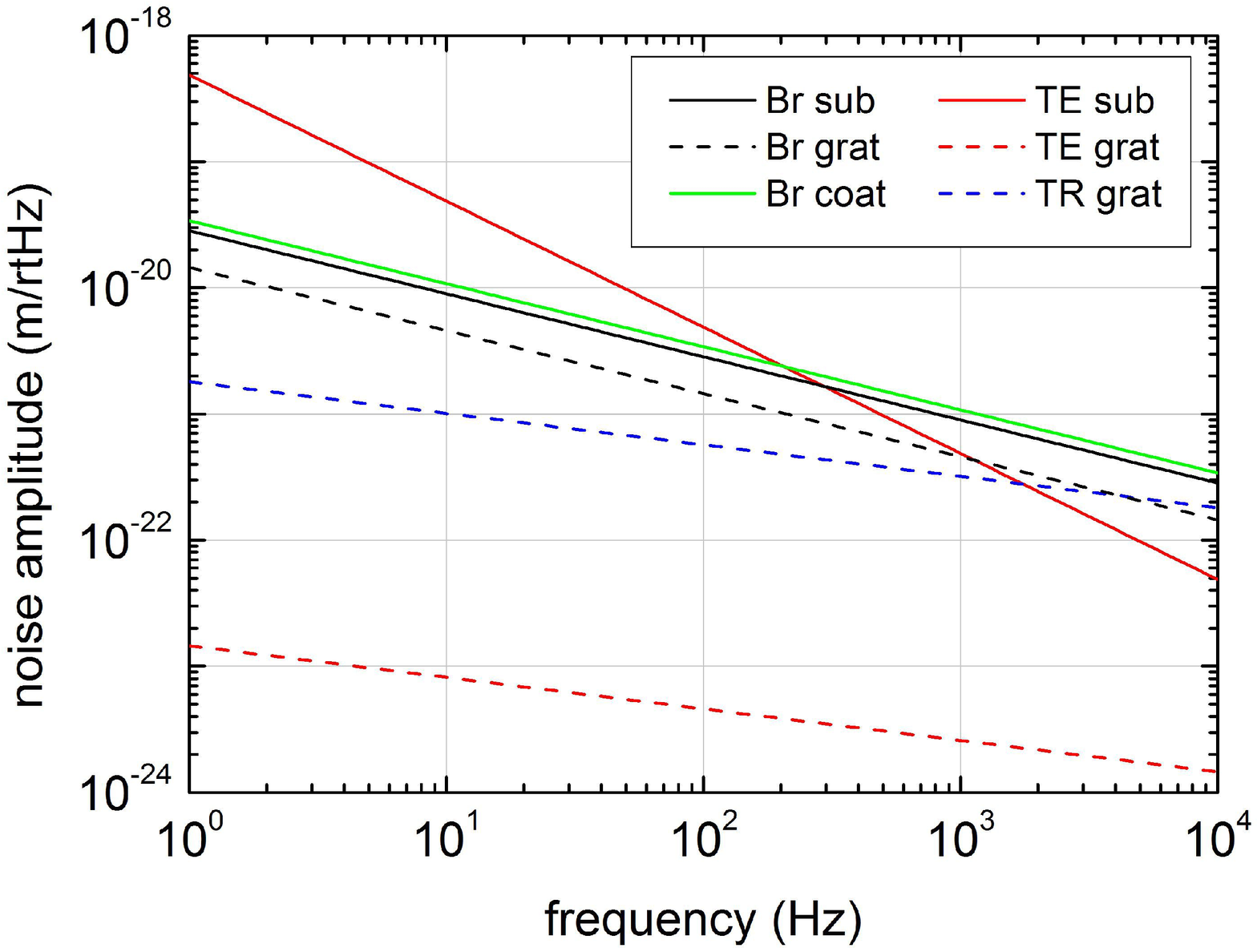}
		\label{fig:Si300K}
  }
  \subfigure[\ $T=\SI{10}{\kelvin}$]{
    \includegraphics[width=7.5cm]{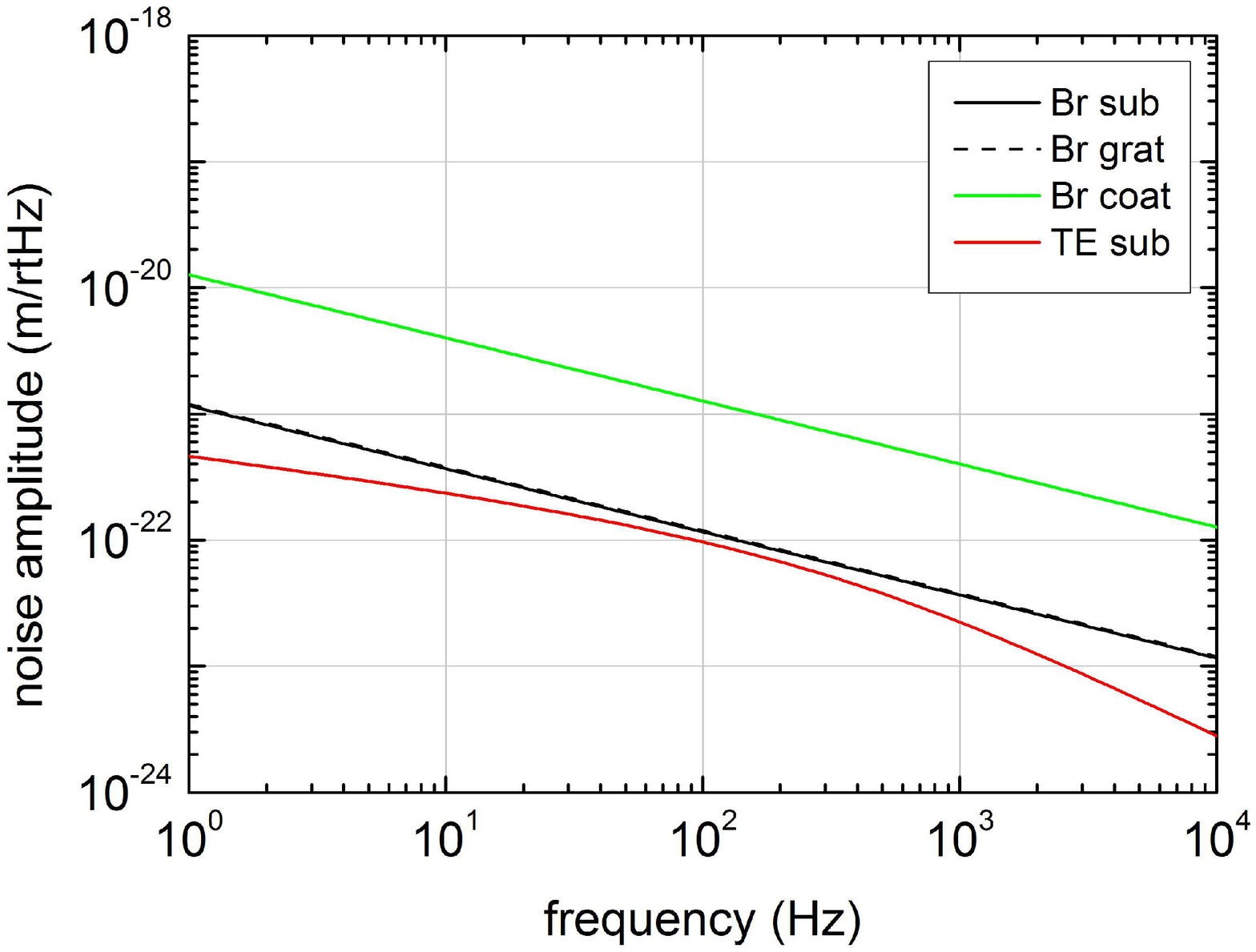}
		\label{fig:Si10K}
  }
	\caption{Thermal noise of a monolithic silicon grating at (a) \SI{300}{\kelvin} and (b) \SI{10}{\kelvin}. At room temperature the TE noise contribution prevents the use of silicon as a substrate material at frequencies below $\sim \SI{500}{\hertz}$. At low temperatures the contributions of TE and TR grating noise drop below the presented noise interval. Brownian noise of a conventional layer stack is significantly above the Brownian noise level of a grating reflector. Note that the grating Brownian noise nearly matches the substrate Brownian contribution at \SI{10}{\kelvin}.  }
\end{figure}

Further the frequency dependence of the grating noise processes has been investigated.
Fig.~\ref{fig:Si300K} and Fig.~\ref{fig:Si10K} show the results for the silicon grating at \SI{300}{\kelvin} and \SI{10}{\kelvin}, respectively.
For the purpose of a convenient comparison the noise performance of conventional layer stacks made of silica and tantala are added to the diagram.
This calculation is based on the theory of Somiya and Yamamoto for finite test masses \cite{somiya2009}.
It assumes a layer stack designed for $\lambda_0=\SI{1550}{\nano\meter}$ showing 19 $\lambda$/4 layers of tantala and silica and an additional $\lambda/4$ silica layer on top of the stack.
With this configuration the stack's reflectivity fulfills the requirements for an aLIGO end mirror.
The underlying material parameters of the coatings are shown in Table~\ref{tab:fusi}.

At first we focus our discussion on the noise contributions of the grating.
The resulting noise spectra clearly indicate, that Brownian noise marks the dominating contribution on the grating performance.
At room temperature the TR contribution is in the same order for frequencies above \SI{1}{\kilo\hertz}, while the TE grating noise is two orders of magnitude below the Brownian contribution.
For cryogenic temperatures with respect to the third law of thermodynamics the thermal expansion $\alpha$ and the thermooptic coefficient $\beta$ approach zero. 
Due to this behavior in silicon both effects become negligibly small at \SI{10}{\kelvin}.
Furthermore, the thermal energy as a driving force for the fluctuations is reduced at low temperatures and promises a second benefit.
Compared to the Brownian noise amplitude of a conventional layer stack our results promise a noise decrease by the factor of 2 and 10 at \SI{300}{\kelvin} and \SI{10}{\kelvin}, respectively.
As a conventional mirror's noise is limited by the coating, the use of grating reflectors significantly improves the noise performance of these devices.

With a decrease of thermal noise of the reflective element the substrate contributions can become crucial and possibly spoil the performance of the mirror.
Thus, the thermal noise contributions of the substrate are added to Figs.~\ref{fig:Si300K} and \ref{fig:Si10K}.
For this purpose the Brownian and TE substrate noise was evaluated by the theories of Liu and Thorne \cite{liu2000} and Cerdonio et al. \cite{cerdonio2001}, respectively.
At room temperature the substrate noise contributions dominate the noise of a silicon sample.
Here the Brownian substrate noise reaches a level equal to Brownian coating noise.
Further at frequencies below \SI{500}{\hertz} TE substrate noise even exceeds the Brownian noise due to the high thermal expansion of crystalline silicon.
Consequently, silicon as a substrate material is not suited for an application at room temperature.

The situation changes for low temperatures.
The decrease in $\alpha$ and $\beta$ leads to an effective suppression of TE substrate noise.
At \SI{10}{\kelvin} TE and Brownian substrate noise as well as Brownian grating noise are roughly a factor of ten below the noise of a conventional layer stack.
In this case the change from a conventional mirror to a grating reflector will effectively reduce the total thermal noise by the same factor of ten.
Consequently, the sensitivity of a low temperature detector made of silicon could be largely increased by using a monolithic grating reflector.
 
\begin{figure}[tb]
	\begin{center}
		\includegraphics[width=7cm]{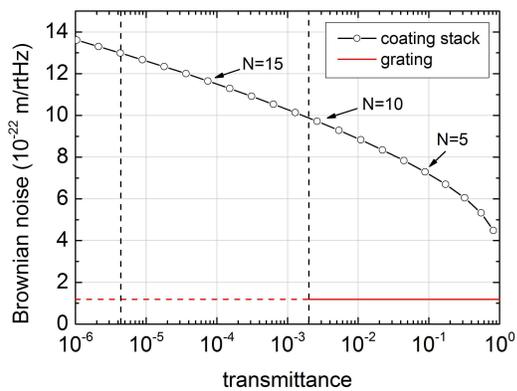}
		\caption{Brownian noise of a conventional silica/tantala layer stack on top of a silicon substrate vs. the transmittance of the layer stack. 
		The diagram shows the noise amplitude with respect to the number $N$ of $\lambda/4$ layer pairs. The results are obtained for a frequency of \SI{100}{\hertz}, a wavelength of \SI{1550}{\nano\meter} and a temperature of \SI{10}{\kelvin}. Also the grating noise is inserted as the horizontal line. Further the dashed lines indicate a stack with N=19 layer pairs (showing the proposed aLIGO reflectivity) and the best measured performance of a T-shaped grating so far.}
		\label{fig:transmission}
	\end{center}
\end{figure}

In addition to the noise spectrum at \SI{10}{\kelvin} shown in Fig.~\ref{fig:Si10K} we varied the number of layer pairs in the calculation of Brownian coating noise.
This treatment allows a comparison taking into account the fact that the reflectivity of current grating reflectors is below that of a Bragg mirror with 19 layer pairs.
The result is presented in Fig.~\ref{fig:transmission}.
There the dashed line at $N=19$ layer pairs again shows the discussed factor of 10 noise improvement.
The highest reflectivity of a T-shaped grating so far is reported as 99.8\%.
This value is illustrated by the second dashed line between $N=10$ and $N=11$ layer pairs.
Even at this reduced reflectivity a grating reflector offers a noise amplitude reduction by a factor of 8.

\subsection{Lamellar tantala grating}
\begin{figure}[tbp]
  \centering
  \subfigure[]{
    \includegraphics[width=7cm]{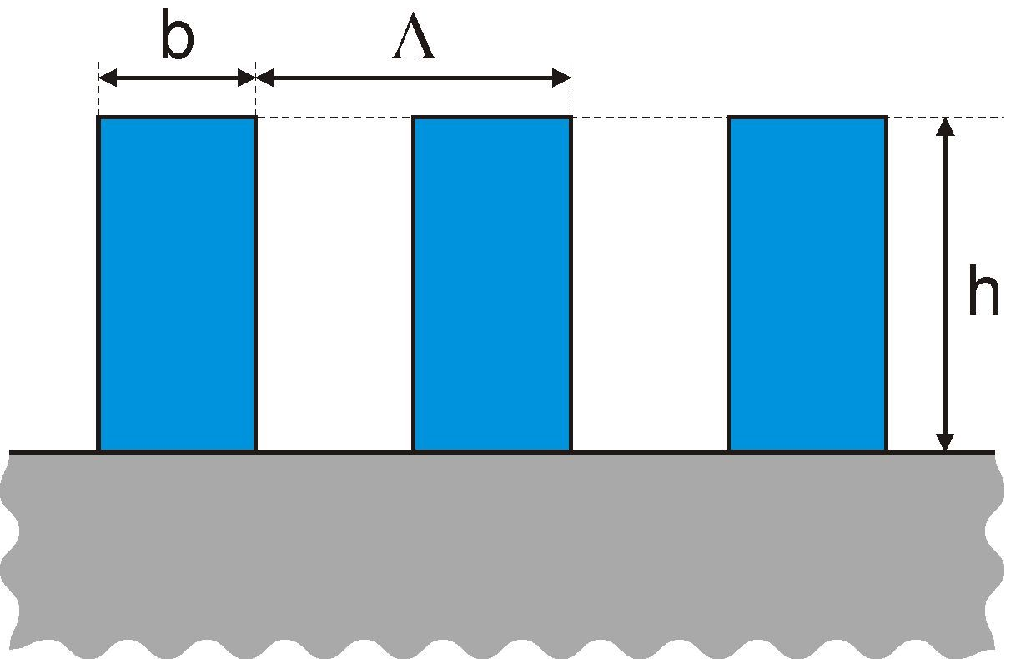}
		\label{fig:tantala}
  }
  \subfigure[]{
    \includegraphics[width=7cm]{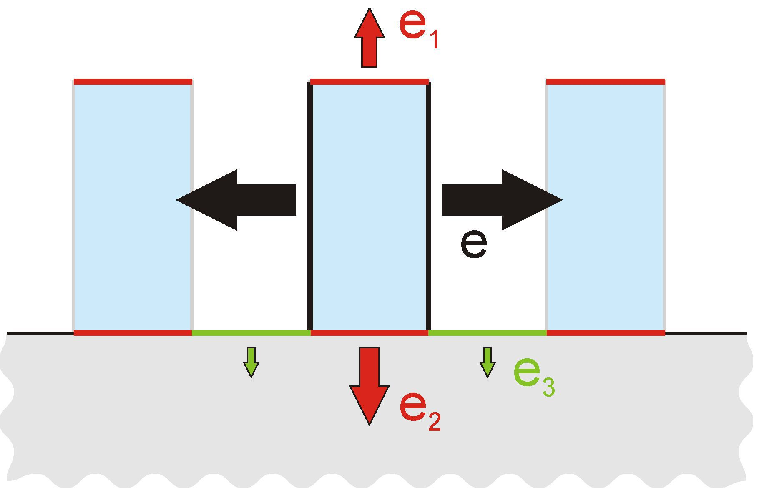}
		\label{fig:tantalaforce}
  }
	\caption{ (a) Alternative highly reflective grating structure. The tantala grating (blue) exhibits two propagating modes and thus allows the manifestation of interferometric effects. The fused silica substrate with its lower refractive index only exhibits a single propagating mode and guarantees a high reflectivity as well as the structural stability of the grating. The respective imaginary forces for a noise analysis are sketched in (b).}
\end{figure}

There exists a second technological realization of highly reflective gratings, featuring a tantala grating on top of a fused silica substrate.
In contrast to silicon, fused silica is known for its low absorption at \SI{1064}{\nano\meter} \cite{hild2006}.
This allows its use in the optical setup of existing gravitational wave detectors. 
The tantala grating incorporating only a single grating region is shown in Fig.~\ref{fig:tantala}.
Here the supporting structure is replaced by the fused silica substrate.
Due to its low refractive index the substrate exhibits only a single propagating mode like the supporting structure in the monolithic grating.
We numerically optimized such a grating to a high reflectivity at $\lambda_0=\SI{1064}{\nano\meter}$ using the RCWA code.
Its parameters are presented in Table~\ref{tab:tantala}.
In this case the substrate is also probed by the evanescent modes, leading to an introduction of noise to the reflected light.

\begin{table}
\caption{Geometry parameters for a single tantala grating on a fused silica substrate. This grating is designed to show a maximum reflectivity at a wavelength of $\lambda_0=\SI{1064}{\nano\meter}$.}
\label{tab:tantala}
\begin{ruledtabular}
\begin{tabular}{ll}
\multicolumn{2}{l}{\textbf{reflective grating}} \\
grating period $\Lambda$& \SI{680}{\nano\metre} \\
refractive index $n$&2.1 \\
depth $h$& \SI{570}{\nano\metre} \\
fill factor $f=b/p$& \SI{.52}{} \\
\addlinespace[3mm]
\multicolumn{2}{l}{\textbf{substrate}}\\
refractive index $n_\mathrm{sub}$ & \SI{1.45}{} \\
\end{tabular}
\end{ruledtabular}
\end{table}

\begin{figure}[tb]
	\begin{center}
		\includegraphics[width=7.5cm]{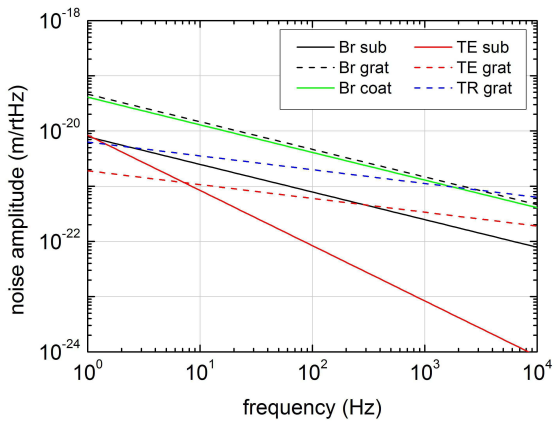}
		\caption{Thermal noise of a tantala grating on a fused silica substrate at \SI{300}{\kelvin}. Brownian grating noise exceeds the level of a conventional layer stack. Compared to a conventional mirror TR grating noise is also likely to limit the noise performance of a tantala grating reflector at frequencies above $\sim\SI{1}{\kilo\hertz}$.}
		\label{fig:fusi300K}
	\end{center}
\end{figure}

Remember that the T-shaped silicon gratings are limited by Brownian noise of the supporting structure.
As a lamellar grating does not show a supporting structure, it promises a decrease in Brownian noise.
Nevertheless a quantitative calculation of noise is sketched below.
Following the load scheme of Fig.~\ref{fig:tantalaforce} a virtual force of $e_1$ has to be applied to the top of the rectangular grating while $e_2$ and $e_3$ are attached to the substrate boundary.
Because grating noise is not affected significantly by the forces to the substrate, $e_2$ and $e_3$, only a value for $e_1$ is important.
An RCWA analysis reveals $e_1=-0.80$.
The force probing a width change possesses an amplitude of $e=5.9$ at each side of the grating.
Again both values have been validated with the analytical model of Ref.~\cite{karagodsky2010} to an accuracy of better than 1\%.
Together with the grating geometry these values prove the width fluctuation to be the dominant one.
In Fig.~\ref{fig:fusi300K} the numerical results of a grating reflector's Brownian noise are summarized and compared to a conventional layer stack optimized for $\lambda_0=\SI{1064}{\nano\meter}$.
Again these results are based on the substrate geometry shown in Table~\ref{tab:substrate}.
The material properties of tantala and fused silica are summarized in Table~\ref{tab:fusi} and Table~\ref{tab:fusisub} of Appendix~\ref{app:matprops}. 
The results show that grating Brownian noise exceeds the noise spectrum of a conventional layer stack as the major noise source in conventional mirrors.
Consequently, the use of tantala gratings is likely to degrade the noise performance compared to multilayer mirrors.

In the case of TR noise we find a coefficient of $K_\mathrm{TR}=21.7$ for a refractive index change in the tantala ridges.
The TR noise contribution from the substrate is characterized by $K_\mathrm{TR,sub}=9.9$.
With our assumed thermooptic parameter of tantala TR noise turns out to be the dominant noise source for frequencies above \SI{3}{\kilo\hertz}.
In the literature there exists a huge variation in the thermooptic coefficient $\beta$ of tantala.
Depending on the actual material TR noise can even exceed the Brownian contributions and totally spoil the noise spectrum of a tantala grating reflector.
In contrast to TR and Brownian noise TE noise appears to be an order of magnitude lower and marks no limitation.
TE noise with respect to the optical grating shows $K_\mathrm{TE,opt}=27.7$.
Also the change of the grating period due to the thermal expansion of the substrate is considered leading to a total value of $K_\mathrm{TE}=32.6$ that was used in the calculation.
For both - TR and TE noise - the temperature noise given in Eq.~(\ref{equ:ST}) was evaluated using the parameters of fused silica as the substrate material.

As the mechanical loss of fused silica increases with decreasing temperature \cite{travasso2007}, cooling can even lead to an increased Brownian substrate noise.
Therefore, cooling does not represent a promising approach to reduce thermal noise in this case. 
For this reason the use of a lamellar tantala grating reflector seems to be no solution to the reduction of thermal noise.
There exists an alternative reflector geometry incorporating an additional tantala layer between grating and substrate.
It is known to exhibit a broader reflectivity spectrum and for that reason promises a lower phase noise.
However, these structures have not been considered in this work.

\section{Conclusion}

With the presented model of thermal noise in grating reflectors we estimated the noise behavior of a monolithic silicon grating. 
At low temperatures the use of a grating reflector promises a decreased thermal noise compared to conventional multilayer mirrors.
Thus, for a typical low temperature third generation gravitational wave detector, the decrease of noise allows a sensitivity increase by a factor of 10.
As we chose a fixed grating configuration in our analysis a systematic design optimization of a grating reflector is likely to even show a higher benefit in terms of thermal noise.
Furthermore, the use of grating reflectors is compatible with other suggested noise reduction methods, e.\,g. the spatial separation of the reflecting planes \cite{khalili2005,somiya2011} or the change of the beam shape \cite{dambrosio2004, mours2006}. 

Due to the obtained noise benefit the Brownian substrate noise will show a comparable impact on the noise performance of a grating reflector.
Only a systematic knowledge of mechanical loss processes in the substrate permits a further improvement in terms of thermal noise in grating reflectors.
This fact strongly calls for a continued experimental investigation into the mechanical loss of proposed substrate materials.

Our work also revealed that an alternative design of a lamellar tantala grating on top of a fused silica substrate shows no benefit in terms of reduced noise compared to a conventional mirror.
Consequently, even if these gratings are easily applicable to existing GW detectors with fused silica substrates at room temperature, a modification of these detectors does not appear worthwhile.
In both cases a systematic optimization of the noise performance of grating reflectors remains as an open task for future investigations.

Currently, experiments on monolithic silicon grating reflectors have demonstrated a maximum reflectivity of 99.8\%.
In the future this value has to be further increased to allow an effective use in the field of frequency stabilization.
Nevertheless their reflectivity already approaches a level required to form an arm cavity in a third generation gravitational wave detector, but this application demands extremely small optical scattering from the mirrors.
So far no precise investigation of the scattering performance has been published.
Experimental characterizations of stray light - especially for monolithic silicon gratings - are therefore highly desirable for an application of grating reflectors in the field of gravitational wave detection.
Further a technological improvement in terms of the grating area is necessary for an effective implementation of grating reflectors into a detector.

\section*{Acknowledgement}
We thank F. Fuchs for providing us with the RCWA code. 
D. H., S. K., R. N., A. T. acknowledge the support by the German Science Foundation (DFG) under contract SFB Transregio 7.
S. P. V. is supported by the Russian Foundation for Basic Research Grants No. 08-02-00580-a and No. 13-02-902441 and NSF grant PHY-0967049.
S. H., S. L, I. W. M. are grateful for the support from the Science and Technologies Facilities Council (STFC). 
S. H. is also grateful for support from the European Research Council (ERC-2012-StG: 307245).
I. W. M. is further supported by a Royal Society University Research Fellowship.
This work has been performed with the support of the European Commission under the Framework Programme 7 (FP7) "People", project ELiTES (Grant Agreement 295153).

\begin{appendix}


\section{Virtual forces in a grating reflector}
\label{app:levinscheme}

In this appendix we will deal with the application of Levin's force to a grating.
For this purpose a simple lamellar grating is used.
Here exemplarily only one degree of freedom is investigated, namely the $z$-displacement of the topmost grating boundary.
The bottom boundary shall be mechanically fixed.

\begin{figure}[b]
	\begin{center}
		\includegraphics[width=7cm]{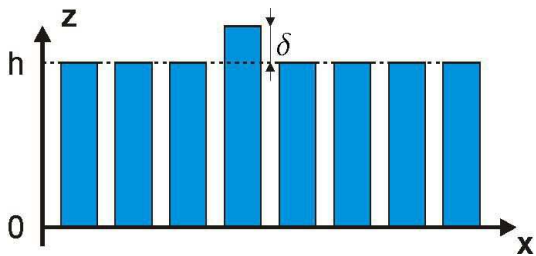}
		\caption{The exemplary lamellar grating with the deformation $\delta$ of a single ridge. The back side of the grating at $z=0$ is mechanically fixed.}
		\label{fig:pertgrat}
	\end{center}
\end{figure}

\begin{figure}[tb]
	\begin{center}
		\includegraphics[width=7.5cm]{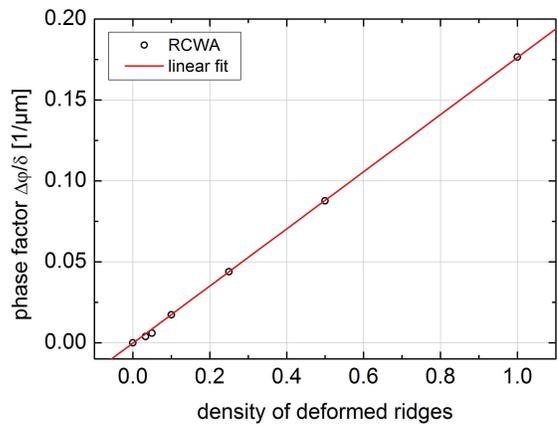}
		\caption{Dependency of the phase slope in the reflected wave with respect to the relative density of the displaced gratings. The linear graph indicates an interaction-free behavior. Thus the motion of each ridge can be treated separately.}
		\label{fig:resdens}
	\end{center}
\end{figure}

Our main interest is in the effect of a single ridge's displacement on the phase of the reflected light.
Fig.~\ref{fig:pertgrat} illustrates this perturbation as well as the model grating.
Typically, however, the RCWA codes only tackle periodic problems.
For this reason we performed a calculation on the behavior of a supercell with 30 periods.
Within them one ridge was displaced as described above.
The high number of ridges in the supercell was chosen to guarantee an interaction-free behavior of the displaced ridges.
Furthermore, we considered an incoming plane wave traveling along the negative $z$-direction.
For this configuration the phase change of the reflected light $\varphi(\delta)$  arising from the displacement $\delta$ was computed in the linear order.
Then the resulting value of $d\varphi/d\delta$ has been computed for a decreasing size of the supercell.
Fig.~\ref{fig:resdens} shows the behavior of this phase slope over the density of the perturbed ridges.
The linear dependence suggests no significant interactions between the deformation of neighbored ridges.
Thus a linear theory on the variation of the reflected light's phase will be applied.
We further checked this idea by the calculation of a randomly displaced supercell with a vanishing mean displacement.
These calculations confirmed an independent treatment.

Before we look for a possible force it shall be noted that only the displacement of the ridges affects the phase of the reflected light.
Thus the grooves of the grating are not probed by the light.
Consequently, a virtual pressure should only be applied to the ridges.
As shown above the effect of a ridge displacement $u_z$ onto the reflected light's phase $\Delta \varphi$ is independent from the displacement of neighboring ridges.
Assuming a homogeneous behavior within a single ridge a generalization of this idea leads to
\begin{align}
\Delta \varphi&=K_0 \frac{\int u_z(x,y)w(x) \exp\left(-\frac{x^2+y^2}{r_0^2} \right)dx dy}{\int w(x) \exp\left(-\frac{x^2+y^2}{r_0^2} \right)dx dy} \\
&=\frac{K_0}{\pi r_0^2 f}\int u_z(x,y)w(x) \exp\left(-\frac{x^2+y^2}{r_0^2} \right)dx dy \ ,
\end{align}
where the integration is performed along the surface plane ($z=h$) of the grating.
Here $w(x)$ equals one at a grating ridge and zero at a groove.
Thus it assures an integration only over the ridges.
Following this model a homogeneous deformation of the grating leads to a phase change of $K_0 u_z$.
A numerical analysis of this case results in a value for $K_0$.
Expressing the phase change in an effective displacement $\Delta z$ directly leads to the derivation of Levin's forces.
For that purpose a virtual pressure of
\begin{align}
p_\mathrm{L}=F_0 \frac{\lambda_0}{4\pi} \frac{K_0}{\pi r_0^2 f}\exp \left(-\frac{x^2+y^2}{r_0^2} \right) \ ,
\label{equ:levinpress}
\end{align}
should be applied to the ridges while no pressure is applied to the grooves.
This procedure is to be adapted for an application to the remaining boundaries in a real grating, which leads to the virtual load scheme presented in Fig.~\ref{fig:load}.

\section{Basics on the plane strain analysis}
\label{app:plainstrain}
\begin{figure}[tb]
	\begin{center}
		\includegraphics[width=5cm]{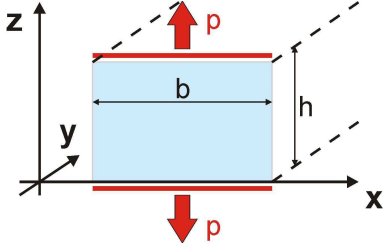}
		\caption{Basic sketch for a homogeneous and symmetric pressure on a bar with a rectangular cross section.}
		\label{fig:sampleload}
	\end{center}
\end{figure}

This appendix will supply the necessary equations to describe the elastic response of a grating ridge.
As a prototype to our direct noise analysis we investigate a homogeneously applied force on both sides of a rectangular cross section.
This load is shown in Fig.~\ref{fig:sampleload}.

A solution to this problem follows from the consideration that the stress component $\sigma_{zz}=p$ is constant within the sample and equals the applied pressure $p$.
Furthermore in a plane strain analysis the strain component $u_{yy}$ has to vanish as it points out of the plane.
As no force is applied pointing along the $x$-direction $\sigma_{xx}=0$ will vanish within the sample.
Using the basic equations for an isotropic elastic solid \cite{LL7} this results in the stress component
\begin{align}
\sigma_{yy}=\sigma \sigma_{zz}=\sigma p \ .
\end{align}
Here $\sigma$ represents Poisson's ratio.
Due to the symmetric load the sample does not exhibit any shear stress components.
Consequently the strain energy density simplifies to
\begin{align}
dw=\sum_{ij}\sigma_{ij}du_{ij}=\sigma_{zz}du_{zz} \ .
\end{align}
Using the elastic equation \cite{LL7}
\begin{align}
u_{zz}=\frac{1}{Y}\left[\sigma_{zz}-\sigma \left(\sigma_{xx}+\sigma_{yy} \right)\right]=\frac{p}{Y}\left(1-\sigma^2\right) \ ,
\end{align}
the strain energy density reads
\begin{align}
w=\frac{1-\sigma^2}{2Y}p^2 \ .
\end{align}

Along the typical ridge dimension the applied virtual pressure in the noise analysis is nearly constant, so the equations above can be used.
Nevertheless, on the larger scale of the beam diameter the pressure shows significant deviations. 
In a simple estimate the total elastic energy can then be calculated by the sum of small parts showing constant pressure.
This results in
\begin{align}
E_\mathrm{tot}=\int \frac{1-\sigma^2}{2Y}p^2(x,y) dV \ ,
\end{align}
where the integration is performed over the grating volume.
By using Eq.~(\ref{equ:levinpress}) in the integral we find
\begin{align}
E_\mathrm{tot}&=\frac{1-\sigma^2}{2Y}\hat{p}^2\int_0^h dz\int_0^b dx\int dy \exp\left(-2\frac{x^2+y^2}{r_0^2} \right) \\
&=\frac{1-\sigma^2}{2Y}\hat{p}^2 h f\frac{\pi r_0^2}{2} \\
&=\frac{1-\sigma^2}{4Y} \frac{h}{\pi r_0^2} \frac{1}{f} \left(K_0\frac{\lambda_0}{4\pi} \right)^2F_0^2 \ .
\label{equ:Etotplanestrain}
\end{align}
Here $\hat{p}$ is introduced as (cf. Eq.~\ref{equ:levinpress})
\begin{align}
\hat{p}=	F_0 \frac{\lambda_0}{4\pi} \frac{K_0}{\pi r_0^2 f}	\ .
\end{align}
The term in brackets can be interpreted as a weighting factor determining the strength of phase noise due to a displacement.

A helpful measure is the linear energy density along the $y$-direction. 
For a single ridge with a rectangular cross section it reads
\begin{align}
E_y=\frac{dE_\mathrm{tot}}{dy}=\int\int w dx dz =bh\frac{1-\sigma^2}{2Y} p^2 \ .
\label{equ:linEdens}
\end{align}
With the typical quadratic dependence on the applied pressure it can further be used to calculate the total elastic energy  of the whole grating by the Gaussian readout as
\begin{align}
E_\mathrm{tot}&=\sum_i\int_{-\infty}^\infty dy E_y(p(x_i,y)) \\
&=\sum_i\int_{-\infty}^\infty dy E_y(\hat{p})\exp\left(-2\frac{x_i^2+y^2}{r_0^2}\right) \\
&=E_y(\hat{p})\sqrt{\frac{\pi}{2} r_0^2} \frac{1}{\Lambda} \: \Lambda \sum_i \exp\left(-2\frac{x_i^2}{r_0^2}\right)
\ .
\end{align}
Here the summation on $i$ is performed over all grating ridges.
As the grating period is small compared to the beam radius the sum is replaced by an integral in the following step
\begin{align}
E_\mathrm{tot}&=E_y(\hat{p})\sqrt{\frac{\pi}{2} r_0^2} \frac{1}{\Lambda} \: \int dx \exp\left(-2\frac{x^2}{r_0^2}\right)  \\
&=\frac{\pi r_0^2}{2\Lambda} E_y(\hat{p}) \ .
\end{align}
Inserting the linear energy density from Eq.~(\ref{equ:linEdens}) leads to the same result for the total energy stored in the grating as shown in Eq.~(\ref{equ:Etotplanestrain}).
Nevertheless the value of an approach using the linear energy density is found in its simple implementation by an FE analysis.
The calculation of $E_y$ is a standard task in FEA for arbitrary shapes and loads and directly leads to the noise behavior.
Therefore it promises an efficient generalization of the above theory on arbitrarily shaped gratings.

Using the COMSOL package \cite{comsol} we firstly checked its results for $E_y$ on the load scheme presented in this appendix.
Later COMSOL was also used to obtain an exact value of $E_y$ for our grating geometry and the load scheme given in Fig.~\ref{fig:load}.

\section{Material properties}
\label{app:matprops}

Table~\ref{tab:siprops} presents the physical properties of silicon used to perform the quantitative estimates of Sec.~\ref{sec:results}.
These parameters are chosen in a way to obtain a pessimistic approach for the grating noise level.
Especially the mechanical loss of the small silicon ridges has been taken from real measurements on microcantilevers (see e.\,g. \cite{yasumura2000, lee2005, mamin2001}).
As these experiments mainly investigate flexural modes, they particularly probe the surface of the samples, which is known to exhibit an increased loss \cite{nawrodt2013}.
In contrast in the grating reflector analysis the sample is probed more homogeneously.
Further the clamping in the experiments is likely to introduce additional loss.
For these reasons the use of the measured loss data represents an upper bound of the real intrinsic values.
Also for the calculation of Brownian substrate noise the bulk loss of silicon was taken from published experiments.
At low temperatures we used a mechanical loss of $\phi=\SI{5e-10}{}$ which has been observed by McGuigan et al. \cite{mcguigan1978}.
The material parameters of fused silica at room temperature are given in Table~\ref{tab:fusisub}.

The properties of silica and tantala coatings are presented in Table~\ref{tab:fusi}.
There the whole parameters have been used for the thermal noise calculation of tantala gratings.
In contrast the calculation of Brownian noise of a layer stack is only affected by the elastic properties $Y$ and $\sigma$ and the mechanical loss $\phi$.
Thus for silica only these parameters are presented in Table~\ref{tab:fusi}.
In our noise estimate we used the loss values of titania-doped tantala that shows decreased losses compared to pure tantala layers.
At \SI{10}{\kelvin} measurements show a further increased loss.
Then also silica reaches a loss level far above \SI{1e-4}{} increasing the resulting noise level of Brownian coating noise.

Recent investigations have identified a layer stack of silica and tantala to exhibit a higher loss than the individual coating layers \cite{granataprep}.
This observation suggests an interaction between the layers changing the mechanical loss of the compound from a linear sum.
The loss values used in this work are based on measurements on single coatings. 
Also the noise estimate of the layer stack has been performed in a linear theory neglecting interactions.
For this reason our result may underestimate the noise level of a conventional layer stack.

\begin{table}
\caption{Material properties of silicon used in this work. In the elastic calculation the silicon substrate and grating are approximated as an effective isotropic medium. Here an orientation along the $\left\langle 100 \right\rangle$ direction was chosen.}
\label{tab:siprops}
\begin{ruledtabular}
\begin{tabular}{lrr}
																&		$T=\SI{300}{\kelvin}$	&	$T=\SI{10}{\kelvin}$ \\
\hline
$\alpha$ [\SI{}{\per\kelvin}]	\cite{white1997}									&\SI{2.62e-6}{}	&\SI{5.0e-10}{} \\
$\beta=dn/dT$ [\SI{}{\per\kelvin}] 	\cite{komma2012}&\SI{1.8e-4}{}		&\SI{1e-7}{} \\
$C$	[\SI{}{\joule\per\kilogram\per\kelvin}]	\cite{bookHull}		&713							&0.276 \\
$\kappa$ [\SI{}{\watt\per\meter\per\kelvin}] \cite{bookTouloukian}&148							&2110 \\

$\rho$ [\SI{}{\kilogram\per\cubic\meter}]				&\multicolumn{2}{c}{\SI{2331}{}}\\
$Y$ [\SI{}{\giga\pascal}]	\cite{wortman1965}		&\multicolumn{2}{c}{\SI{130}{}}\\
$\sigma$ \cite{wortman1965} 										&\multicolumn{2}{c}{\SI{0.28}{}}\\

$\phi_{grat}$																		&	\SI{5e-5}{}	\cite{yasumura2000}		&\SI{1e-5}{} \cite{mamin2001}\\
$\phi_{bulk}$	\cite{mcguigan1978}																	&	\SI{1e-8}{}			&\SI{5e-10}{} \\
\end{tabular}
\end{ruledtabular}
\end{table}

\begin{table}
\caption{Material properties of fused silica as a substrate material at \SI{300}{\kelvin}. Except where noted, material properties are taken from Fejer et al. \cite{fejer2004}.}
\label{tab:fusisub}
\begin{ruledtabular}
\begin{tabular}{lr}
	&		fused silica	\\
\hline
$\alpha$ [\SI{}{\per\kelvin}]								&\SI{0.51e-6}{}	 \\
$n$																							&1.45			\cite{benthem2009}	\\
$\beta=dn/dT$ [\SI{}{\per\kelvin}] 					&\SI{8.5e-6}{}\cite{benthem2009}	  \\
$C$	[\SI{}{\joule\per\kilogram\per\kelvin}]			&746						 \\
$\kappa$ [\SI{}{\watt\per\meter\per\kelvin}]		&1.38						 \\
$\rho$ [\SI{}{\kilogram\per\cubic\meter}]				&2200						 \\
$Y$	[\SI{}{\giga\pascal}]												&72							 \\
$\sigma$																				&0.17						 \\
$\phi$																					&\SI{4e-10}{}	\cite{penn2006}	 
\end{tabular}
\end{ruledtabular}
\end{table}

\begin{table}
\caption{Material properties of silica and tantala coatings at \SI{300}{\kelvin}. Except where noted, material properties are taken from Evans et al. \cite{evans2008}. For silica only parameters affecting Brownian noise are given.}
\label{tab:fusi}
\begin{ruledtabular}
\begin{tabular}{lrr}
	&		tantala	&	silica \\
\hline
$\alpha$ [\SI{}{\per\kelvin}]								&\SI{3.6e-6}{}				&\SI{0.51e-6}{} \\

$\beta=dn/dT$ [\SI{}{\per\kelvin}] 					&\SI{14e-6}{}				&  \\
$\rho$ [\SI{}{\kilogram\per\cubic\meter}]				& 6850			& \\
$Y$	[\SI{}{\giga\pascal}]												&140				&72 \\
$\sigma$																				&0.23				&0.17 \\
$\phi$ at \SI{300}{\kelvin}											&\SI{2.4e-4}{}\cite{flaminio2010}	&\SI{5e-5}{} \cite{penn2003} \\
$\phi$ at \SI{10}{\kelvin}						&\SI{4e-4}{} \cite{martin2008}		&\SI{4.5e-4}{} \cite{phDmartin}\\
$n$																							&2.07 \cite{flaminio2010}				&1.45\\
\end{tabular}
\end{ruledtabular}
\end{table}

\section{Mechanical energy in grating ridges}
\label{app:bending}
In this appendix we illustrate the solution of the elastic equations for the grating reflector under the virtual loads.
For this purpose a lamellar grating on top of a substrate is used and our approach is restricted to elastically isotropic materials.
A virtual pressure is then to be applied to the top of the grating ridges
\begin{align}
\sigma_{zz}=\frac{1}{f}\frac{F_0}{\pi r_0^2}\exp\left(-\frac{r^2}{r_0^2} \right) \ .
\label{equ:virtszz}
\end{align}
Here the normalization by the fill factor $f$ ensures a constant amplitude of the virtual force for gratings of arbitrary geometries.

We further follow the considerations of Harry et al. \cite{harry2002}.
Due to the small height of the grating, the response of the reflector to the virtual force is dominated by the substrate.
The elastic energy of the ridges is then obtained via transition conditions.
There we adopt the treatment of a coating, where $\sigma_{zz}$, $u_{xx}$ and $u_{yy}$ show a continuous behavior along the boundary between the substrate and grating.
In contrast to a coating a grating is not continuous. 
Consequently the free sides should be traction free leading to $\sigma_{xx}=0$ (see Fig.~\ref{fig:lamellar}).
As the ridges show a small thickness, the above stress component has to vanish inside the grating ridge.
Thus the condition $\sigma_{xx}=0$ instead of the substrate value for $u_{xx}$ is applied.
Using the elastic equations \cite{LL7}
\begin{align}
u_{xx}=\frac{1}{Y}\left[\sigma_{xx}-\sigma\left(\sigma_{yy}+\sigma_{zz} \right) \right] \ , \nonumber \\
u_{yy}=\frac{1}{Y}\left[\sigma_{yy}-\sigma\left(\sigma_{xx}+\sigma_{zz} \right) \right] \ , \\
u_{zz}=\frac{1}{Y}\left[\sigma_{zz}-\sigma\left(\sigma_{xx}+\sigma_{yy} \right) \right] \ , \nonumber
\end{align}
and $\sigma_{xx}=0$ we find the following stresses and strains in the grating
\begin{align}
\sigma_{yy}&=Y_g u_{yy}^s+\sigma_g \sigma_{zz} \ , \nonumber \\
u_{xx}&=-\frac{\sigma_g}{Y_g}\left[Y_g u_{yy}^s+(1+\sigma_g)\sigma_{zz} \right] \ , \label{equ:elast}\\
u_{zz}&=\frac{1-\sigma_g^2}{Y_g} \sigma_{zz}-\sigma_g u_{yy}^s \ . \nonumber
\end{align}
Here Young's modulus $Y_g$ and Poisson's ratio $\sigma_g$ of the grating material are introduced.
The strain component $u_{yy}$ has been replaced by the substrate field $u_{yy}^s$ due to the transition conditions.
Furthermore, the stress component $\sigma_{zz}$ has to be inserted from Eq.~(\ref{equ:virtszz}).

We now calculate the elastic energy density of the grating ridges via
\begin{align}
\rho=\frac{\sigma_{zz}u_{zz}}{2}+\frac{\sigma_{yy}u_{yy}}{2} \ .
\end{align}
Here we neglect the shear parts of the energy which are supposed to be small due to the bending load case. 
Using the expressions in Eqs.~(\ref{equ:elast}) we find
\begin{align}
\rho=\frac{1-\sigma_g^2}{2Y_g}\sigma_{zz}^2+\frac{Y_g}{2} \left(u_{yy}^s\right)^2=\rho_1+\rho_2 \ .
\end{align}
The above equation defines $\rho_1$ and $\rho_2$ as the two energy contributions.

In a last preparation step the substrate strain $u_{yy}^s$ has to be evaluated.
For this purpose the substrate solution for a continuous Gaussian load is used.
This approach seems reasonable due to the high ridge density and the small scale of the ridges and grooves compared to the laser beam radius.
Thus the substrate is likely to average the load at the surface between ridges and grooves.
Following Eq.~(A7) in Ref.~\cite{harry2002} the displacement in radial direction reads
\begin{align}
u_r^s=-F_0 \frac{(1+\sigma)(1-2\sigma)}{2\pi Y r}\left[1-\exp\left(-\frac{r^2}{r_0^2} \right)\right] \ .
\end{align}
Here $Y$ and $\sigma$ characterize the elastic properties of the substrate.
Please note that due to the averaged load the fill factor $f$ does not enter the last equation.
The strain component in Cartesian coordinates $u_{yy}^s$ is obtained as
\begin{align}
u_{yy}^s	&=\frac{u_r^s}{r}+\frac{y^2}{r}\partial_r\left(\frac{u_r^s}{r}\right) \\
					&=F_0 \frac{(1+\sigma)(1-2\sigma)}{2\pi Y}\left[\frac{1-\Gamma}{r^2}\left(\frac{2y^2}{r^2}-1 \right)-\frac{2y^2\Gamma}{r^2r_0^2} \right] \ ,
\end{align}
where $\Gamma\equiv\exp(-r^2/r_0^2)$.

Finally the elastic energy of the ridges is obtained by an integration of the energy density and reveals
\begin{align}
E_1&=\int\rho_1 dV=hf\int_0^\infty \rho_1(r) \, 2\pi r dr\\
&=\frac{h}{\pi r_0^2} \frac{1-\sigma_g^2}{4Y_g} \frac{1}{f} F_0^2 \ .
\end{align}
This contribution is identical to the expression of Eq.~(\ref{equ:Etot}).
The second energy contribution reads
\begin{align}
E_2&=\int\rho_2 dV=hf\int_0^\infty\int_0^{2\pi} \rho_2(r,\varphi) \, rd\varphi dr \\
&=\frac{h}{\pi r_0^2} \frac{3(1+\sigma)^2(1-2\sigma)^2}{32Y} \frac{Y_g}{Y} f F_0^2 \ .
\end{align}
Note that this second energy contribution shows a different dependence on the fill factor.
The ratio between both contributions also depends on the fill factor.
For the special case of identical grating and substrate materials ($Y=Y_g$, $\sigma=\sigma_g)$ it reads
\begin{align}
\frac{E_2}{E_1}=\frac{1}{8} \, \frac{1+\sigma}{1-\sigma}(1-2\sigma)^2  f^2 \leq\frac{1}{8}\ .
\end{align}
Thus, the second contribution is smaller than the first one by at least a factor of 8.
This ratio is further increased for an increasing value of Poisson's ratio $\sigma$ or the fill factor $f$.
In typical gratings ($f=0.5$, $\sigma=0.15$) the ratio is 50.
This is why $E_2$ can be neglected in most cases and is also neglected within this work.

The validity of the presented equations has been checked in a numerical 3-dimensional calculation using the FE package COMSOL.
There a macroscopic, monolithic lamellar grating on top of a silicon substrate with ET-LF geometry (see Table~\ref{tab:substrate}) has been investigated.
The grating ridges exhibited a quadratic cross section with a length of \SI{2}{\centi\meter} ($b=h=\SI{2}{\centi\meter}$, $\Lambda=\SI{4}{\centi\meter}$, $f=0.5$).
This macroscopic grating was chosen to allow for a sufficient mesh density in the numerical calculation.
The analytical solution for the energy of the above grating revealed
\begin{align}
E_1&=\SI{5.57e-12}{\joule} \ , &E_2=\SI{0.18e-12}{\joule} \ .
\end{align}
Numerically we found a value of $E_{tot}=\SI{5.6e-12}{\joule}$ representing a deviation of less than 3\% and a clear confirmation of the analytical model.
With this successful numerical test of our theory for a macroscopic grating, a coincidence at smaller scales is also likely as the approximations used are even more valid at small scales.

\end{appendix}
\bibliography{grating}

\end{document}